\documentclass[lettersize,journal]{IEEEtran}
\usepackage{amsmath,amsfonts}
\usepackage{algorithmic}
\usepackage{algorithm}
\usepackage{array}
\usepackage{textcomp}
\usepackage{url}
\usepackage{verbatim}
\usepackage{graphicx}
\usepackage{subcaption}
\usepackage{cite}
\usepackage{svg}
\usepackage{colortbl}
\usepackage{graphicx}
\usepackage{multirow}
\usepackage[normalem]{ulem}
\usepackage{booktabs}
\useunder{\uline}{\ul}{}
\usepackage{mathtools}
\usepackage{hyperref}
\usepackage{makecell}
\usepackage{tikz}
\usepackage{listings}
\usepackage{tablefootnote}
\usepackage{amssymb}% http://ctan.org/pkg/amssymb
\usepackage{pifont}% http://ctan.org/pkg/pifont
\usepackage{dblfloatfix}
\usepackage{array}
\usepackage{float}
\usepackage{wasysym}
\usepackage{xcolor}
\usepackage{caption}
\usepackage{booktabs}
\usepackage{multirow}
\usepackage{float} % Required for the [H] placement specifier

\newcommand{\tool}[1]{\texttt{DAST}}
\newcommand{\updatecolor}{\color{black}}
\newcommand{\update}[1]{{\updatecolor #1}}

\usepackage{eso-pic}

\begin{document}

\title{\texttt{DAST}: A VLM–LLM Framework for Cross-Interface Anomaly Detection in O-RAN}

\author{Francesco Spinelli, Esteban Municio, Pau Baguer, Gines Garcia-Aviles, Xavier Costa-Perez

\thanks{Francesco Spinelli, Esteban Municio, Pau Baguer and Gines Garcia-Aviles are with i2CAT Foundation}
\thanks{Xavier Costa-Perez is with i2CAT Foundation, NEC Laboratories Europe and ICREA}

}

% \markboth{IEEE Network Magazine, Special Issue on Connected and Collaborative IoT--Edge--Cloud Continuum for 6G Networks}%
% {Anonymous \MakeLowercase{\textit{et al.}}: Agentic Vision--Language Reasoning for Anomaly Detection in the 6G Continuum}

\maketitle

\begin{abstract}

O-RAN enables a disaggregated baseband stack with programmable functions that communicate over standardized open interfaces. The same openness that enables multi-vendor composition also expands the attack surface across logically decoupled tiers that make up the compute continuum. Among these threats, Denial-of-Service and performance-degradation attacks, which account for the majority of catalogued O-RAN threats, are particularly difficult to detect. Traditional Time-Series Anomaly Detection (TSAD) methods fail in this new regime where labelled baselines are scarce, threats evolve faster than detectors can be retrained, and the high-dimensional multivariate telemetry overwhelms monolithic inference models. To address these challenges, we present \tool{}, a zero-shot multi-agent framework for cross-interface anomaly detection in O-RAN that chains a three-stage VLM $\rightarrow$ LLM $\rightarrow$ VLM pipeline.
\tool{} converts multivariate KPI streams into visual representations, scores textual per-interface descriptions against O-RAN Domain Knowledge, and verifies suspects on high-resolution heatmaps to output the problematic interfaces, the anomalous time intervals, an indicative O-RAN WG11-aligned operational impact rating and the decision rationale. We evaluate \tool{} on real network traces collected from an O-RAN testbed under representative performance degradation scenarios, achieving $0.911$ F1-Score and $0.844$ Accuracy, outperforming state-of-the-art TSAD baselines.

\end{abstract}

\begin{IEEEkeywords}
LLMs, VLMs, Multivariate anomaly detection, O-RAN, Multi-Agent Systems
\end{IEEEkeywords}

\section{Introduction}
\label{sec:intro}
The transition to sixth-generation (6G) mobile networks is being conceived not only as a faster air interface but as a distributed compute fabric in which connectivity, computation and control are co-designed across a device-edge-cloud continuum. Within this continuum, the Radio Access Network (RAN) ceases to be a monolithic appliance and becomes a set of virtualized functions that compete for, and collaborate over, shared heterogeneous compute resources. Observability of this fabric is no longer a per-box concern; rather, faults or attacks can surface several tiers or interfaces away from their origin.

Driven by this shift, the RAN is evolving into a 
software-driven, programmable layer of a distributed edge-cloud system in which RAN 
functions co-exist with other virtual functions within edge servers and centralized cloud 
resources. This architectural convergence is enabled by the Open RAN (O-RAN) paradigm, 
which disaggregates the baseband stack into functional units (i.e., the O-CU, O-DU and O-RU)
and includes RAN Intelligent Controllers (RICs), all interconnected through open, 
standardized interfaces (e.g., E2, A1, F1-c, F1-u) and deployable across 
heterogeneous compute tiers. However, the same openness and disaggregation that unlock 
the flexibility of 6G also create a complex operational problem. 
Multi-vendor disaggregation exposes every open interface as a potential entry point, 
and third-party xApps/rApps inject external code directly into the radio control loop, 
enabling adversaries to perform signaling storms and stealthy Denial-of-Service 
(DoS) attacks. Detecting them is of utmost importance, since 60\% of all identified 
O-RAN threats correspond to DoS and performance degradation attacks, whose effects cascade across the RAN without triggering  coarse-grained alarms~\cite{oran_wg11_threat_2022}.

Traditional approaches to RAN anomaly detection rely on Time-Series Anomaly Detection (TSAD) methods such as supervised and unsupervised Machine Learning (ML) models. While they perform well on clean, stationary benchmarks (e.g.,~\cite{zhang2019mscred,sun2024spotlight}),
they can sharply degrade when faced with evolving threats and 
network architectures.  
Labelled ground-truth data is scarce and prohibitively expensive to collect at scale across multi-vendor, dynamically configured deployments.  
Furthermore, evaluating each interface in isolation leaves these detectors blind to the cross-interface cascades driven by O-RAN's closed control loops. 
For instance,  a subtle latency degradation on F1-c may later evolve into an F1-u throughput collapse due to attachment failure. These constraints make supervised/unsupervised ML ill-suited to the compute continuum regime where large volumes of unlabelled, non-stationary telemetry data are available.

Applying Foundation Models such as Large Language Models (LLMs) and Vision Language Models (VLMs) directly to KPI telemetry partially avoids the labelling problem~\cite{liu2025llmad,He_Alnegheimish_Reimherr_2026}. However, deploying a single LLM or VLM as the sole inference engine reintroduces the same dimensionality bottleneck, inheriting the well-documented weakness of current LLMs in reasoning over high-dimensional numerical data~\cite{xu2026can}, and treating each series in isolation and ignoring the semantics of the interfaces being monitored~\cite{He_Alnegheimish_Reimherr_2026}.

In this article, we show the potential of combining Foundation Models into a prompt-grounded reasoning pipeline leveraging~\emph{(i)} visual perception,~\emph{(ii)} language reasoning, and~\emph{(iii)} O-RAN Domain Knowledge the way a human network expert would do \cite{sre_book_troubleshooting}. To this end, we present \texttt{DAST} (\textit{Detecting Anomalies and Security Threats in the RAN}), a Network Intelligence (NI) solution designed within the scope of the Compute Continuum Layer (CCL)~\cite{chatzieleftheriou2024towards} for detecting anomalies in O-RAN. \texttt{DAST} is a \textit{multi-modal reasoning pipeline} where VLMs and LLMs operate as collaborative diagnostic agents. It has a three-stage VLM~$\rightarrow$~LLM~$\rightarrow$~VLM chain that mimics the reasoning of a human network expert: a first VLM ingests the plot of each time series coming from O-RAN interfaces, visually profiling them. Then, an O-RAN-grounded LLM compares the profile to the expected behaviour, giving a score. Finally, a second VLM verifies suspect time series on high-resolution heatmaps. The pipeline outputs, as machine-readable report, the exact time intervals of detected anomalies, the involved interfaces, an \update{indicative} operational \textit{impact} rating, and a chain-of-thought rationale, which is intended to feed downstream root-cause analysis and SRE-style troubleshooting~\cite{sre_book_troubleshooting}. 
Decomposing the task across specialised agents bypasses the dimensionality and numerical-reasoning bottlenecks of any single foundation model, while explicit domain grounding replaces the need for retraining: the pipeline is fully zero-shot, requires no fine-tuning or clean baseline datasets, and ingests O-RAN Domain Knowledge that can be easily updated as the specifications evolve.

The main contributions of this work are:
\begin{itemize}
    % \item We present \tool{}, which, to the best of our knowledge, represents the first zero-shot multi-agent VLM-LLM architecture for cross-interface anomaly detection in O-RAN. \tool{} is designed to generalise across multi-vendor configurations and zero-day performance degradation patterns without labelled data or fine-tuning. 

    \item\update{We present \tool{}, which, to the best of our knowledge, represents the first zero-shot multi-agent VLM-LLM architecture for cross-interface anomaly detection in O-RAN. By eliminating the need for labelled data or fine-tuning, \tool{} avoids the retraining bottlenecks typically associated with changing vendor configurations or previously unseen performance degradation patterns.}

    \item We evaluate \tool{} on real network traces obtained from an open-source O-RAN testbed under representative performance degradation scenarios.
    \item We compare \tool{} against several well-known state-of-the-art anomaly detection benchmarks. 
\end{itemize}

\section{Background and Related Work}
\label{sec:background}

\subsection{O-RAN Disaggregation and Attack Surface}
The Open RAN architecture promoted by the O-RAN Alliance decomposes the 
gNB and exposes its internal interactions through a set of open, 
standardized interfaces, with control delegated to two RICs. The Near-RT RIC hosts \textit{xApps} on control loops 
of $10$--$1000$\,ms over the E2 interface, while the Non-RT RIC hosts 
\textit{rApps} on loops above $1$\,s and pushes policies to the Near-RT 
RIC through the A1 interface. Operations and maintenance flows are 
carried over O1, while the O-CU/O-DU/O-RU split is realized over the 3GPP F1-c/F1-u interfaces and Open Fronthaul, typically using eCPRI.

While this openness and modularity enables multi-vendor RAN composition and control, it also enlarges the exposed surface: each open interface becomes an externally reachable boundary, third-party rApps/xApps inject external code into the RAN control loop, and the underlying O-Cloud introduces infrastructure-level attack vectors~\cite{baguer2024attacking}. 

Among these events, many are stealthy by design, converting moderate latency or packet-loss injection on A1, E2 or F1-c/F1-u into degradation of end-user QoE without raising threshold alarms. These types of subtle DoS attacks are notably difficult to detect since they may not originate in the actual O-RAN components, interfaces often do not show clear signs of malfunctioning, and anomalies in one interface can degrade other interfaces, propagating through O-RAN's closed control loops. For example, a signalling-storm or DoS attack initiated on E2 typically cascades onto F1-c and F1-u within seconds, and even moderate packet delays on F1-c can terminate the underlying SCTP association and detach UEs from the network, as well as disrupt attaching procedures~\cite{baguer2024attacking}. 
Reliably detecting such events therefore requires correlated 
observation across multiple interfaces simultaneously and a deep understanding of O-RAN architecture.

\subsection{From Traditional TSAD to Foundation Models}
\label{subsec:related_tsad}

Most traditional TSAD approaches rely on uncontaminated traces and flag deviations at inference.
For instance, \textit{MSCRED} couples convolutional encoders with recurrent decoders to detect anomalies and localise root causes in multivariate streams ~\cite{zhang2019mscred}, while \textit{SpotLight} delivers explainable detection on a carrier-grade O-RAN testbed~\cite{sun2024spotlight}. 
Despite strong in-distribution performance, applying traditional TSAD approaches faces three structural limitations in the O-RAN regime:

\noindent\textbf{(L1)~\emph{Labelled ground truth is scarce}}: 
uncontaminated traces are prohibitively expensive to acquire and keep updated, especially across the combinatorial space of vendors, slices and topologies that real deployments span. Furthermore, zero-day attacks causing performance degradation by definition have no previous examples. 

\noindent\textbf{(L2)~\emph{Distribution shift overwhelms retraining}}: 
multi-vendor integration and dynamic slice reconfiguration produce non-stationary telemetry on a timescale shorter than any feasible retraining cycle. Thus, accuracy
degrades substantially over time~\cite{pendlebury2019tesseract}.

\noindent\textbf{(L3)~\emph{Per-interface detectors are blind to cascades}}: 
Traditional TSAD approaches evaluate O-RAN interfaces in isolation, treating telemetry as decoupled systems. This architecture is blind to the cross-interface cascade effects driven by O-RAN's closed control loops. Lacking a structural and temporal understanding, isolated detectors treat propagating symptoms as disconnected events, flooding operators with a storm of alerts.

A way to bypass these constraints could be to leverage Foundation Models. Indeed, LLMs and VLMs are pre-trained on internet-scale text and image corpora data and present remarkable zero-shot and few-shot capabilities (i.e., they can provide anomaly detection even if not trained on that specific task). It is possible to pretrain LLMs to perform anomaly detection; however these steps generally require a substantial compute infrastructure.
Instead, other approaches leverage the innate natural language processing capabilities of Foundation Models, directly prompting numerical sequences into LLMs~\cite{liu2025llmad}, rendering time series images (e.g., line plots) and sending them to a VLM~\cite{He_Alnegheimish_Reimherr_2026,zhuang2024itthinkitsorted}, or creating a flexible multi-agent system to detect different types of anomalies~\cite{xu2026can}. 

ieee-network-RESUBMISSION/
\begin{figure*}[ht]
  \centering
  \includegraphics[width=\linewidth]{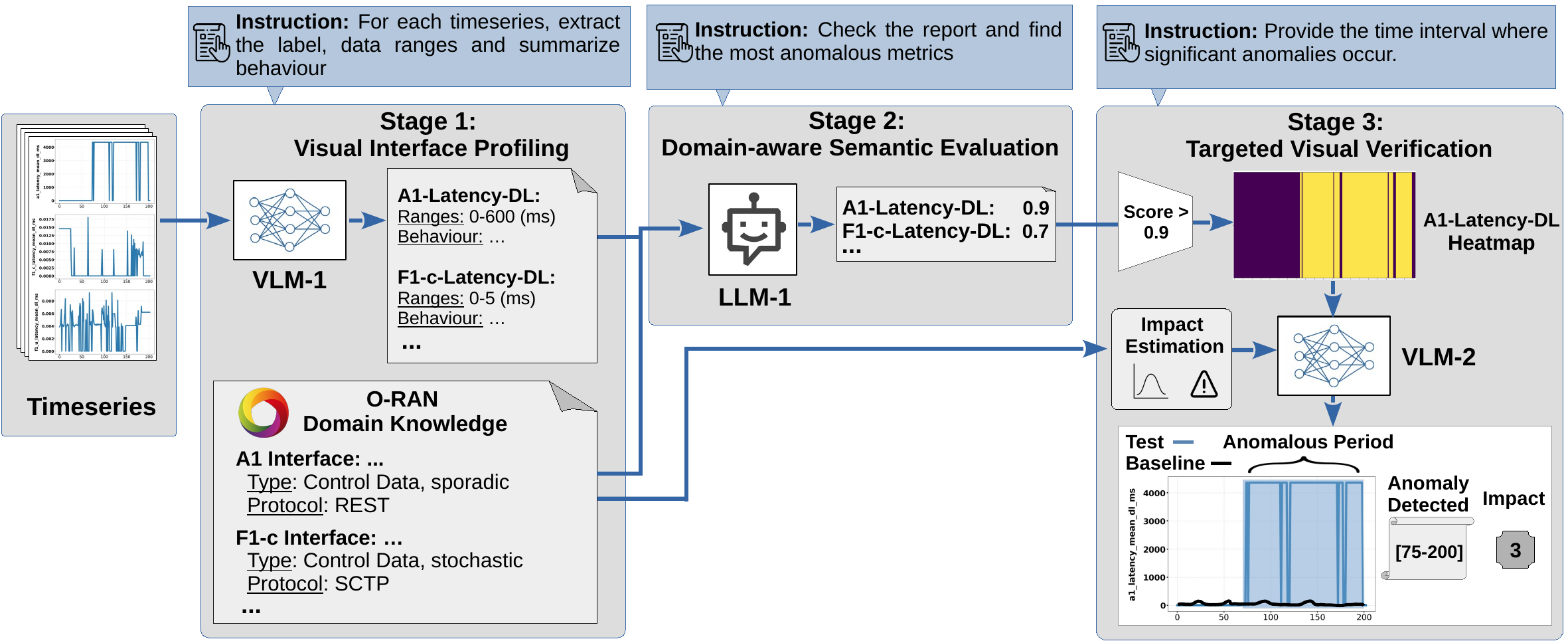}
  \caption{\tool{} architecture. KPI streams from the four O-RAN interfaces are rendered into stacked line plots (Stage 1) and described in text by VLM-1. An O-RAN-grounded LLM (Stage 2) scores per-interface descriptions against expected behaviour. A second VLM (Stage 3) verifies high-scoring suspects on per-metric heatmaps and emits time intervals plus a WG11-aligned impact rating.}
  \label{fig:dast_framework}
\end{figure*}

Within the O-RAN domain, at the time of writing, papers use Foundation Models just as post-hoc explanations on top of trained classifiers~\cite{chatzimiltis2026aionran} rather than as anomaly detectors. However, leveraging a single LLM/VLM to process O-RAN telemetry is not feasible, since concatenating high-rate telemetry from all O-RAN interfaces could exceed the effective context window of current models~\emph{(i)} forcing lossy aggregation and/or~\emph{(ii)} degrading LLM performance due to the large multi-dimensional numerical arrays.

\section{\tool{}: Detecting anomalies in O-RAN enabled 5G networks}
\label{sec:system}

When a seasoned network engineer is handed a wall of O-RAN telemetry, they do not scrutinize every counter against a learned baseline. They first check the plots to sense which signals look unusual, then reason about whether that visual impression is genuinely abnormal given what each interface is supposed to do. \tool{} mirrors this workflow directly. Instead of forcing a single model to do everything, it distributes the task across three specialized, collaborative agents (visual observer, domain-grounded reasoner and fine-grained visual verifier), each of them operating on the metrics representation it handles best, moving from global inspection to interval-level localization (Fig.~\ref{fig:dast_framework}). The agents are fully zero-shot (no training on O-RAN traces), and all O-RAN specific knowledge is supplied as an external, updatable domain-knowledge artefact rather than as labelled examples. The workflow operates through the following stages:\newline

\noindent \textbf{Stage 1: Visual Interface Profiling (VLM).} The pipeline begins by ingesting the complete multivariate time series representing  raw metric data from different O-RAN network interfaces (e.g., UL latency of E2, DL throughput of F1-c, DL packet loss of A1, etc.), and rendering a single, vertically stacked line-plot image to preserve temporal alignment across all channels. A VLM (VLM-1 in Fig.~\ref{fig:dast_framework}) acts as the first-pass observer. It generates a report highlighting the metric label, the axis ranges, and the general behaviour (e.g., steady from $t=0$ to $t=70$, then bursty until $t=120$, finally oscillations until the end of the time window) of each time series. This translates dense visual signals into a structured, explainable textual summary. In parallel, we prepare the O-RAN Domain Knowledge. It includes comprehensive information about interfaces in O-RAN networks such as their roles (e.g., control plane or user plane), what type of traffic they usually have (e.g., light sporadic, heavy stochastic, periodic), what protocols are used by each interface, etc.\newline

\noindent\textbf{Stage 2: Domain-Aware Semantic Evaluation (LLM).} Rather than performing a direct, pixel-by-pixel visual study of the raw time-series and directly trying to detect the anomalies, an LLM ingests the text reports generated in the first stage (LLM-1 in Fig.~\ref{fig:dast_framework}), to decide which time series present the most remarkable anomalies. To help with this decision and narrow the search space, we leverage O-RAN Domain Knowledge to help the LLM reasoning agent. By grounding the system with explicit O-RAN knowledge, and detailing the expected behaviours of O-RAN interfaces (e.g., the high-volume stochastic nature of F1-u versus the strict reliability bounds of F1-c), the LLM can now cross-reference the VLM's visual observations expressed as text against this domain knowledge. As a result, the LLM identifies the most suspicious metrics and assigns each of them a score requiring no labelled training data.\newline
  
\noindent \textbf{Stage 3: Targeted Visual Verification (VLM).} Then, the system applies a filtering threshold. For any metric flagged by the LLM with a high certainty score (i.e., $> 0.9$), the framework dynamically retrieves the heatmap figures of the selected metrics. These heatmap images of the suspicious time series, along with the O-RAN Domain Knowledge, will be used to query the VLM again (VLM-2 in Fig.~\ref{fig:dast_framework}). This second VLM pinpoints the exact time interval where anomalies happen, taking into account possible cascade effects between time series. Finally, the VLM outputs the precise boundaries of the estimated anomaly time intervals. Besides the anomalous time intervals, VLM-2 includes an \textit{impact} rating and a text description of the \textit{thinking} process in its output. The impact rating roughly estimates how critical the anomaly is, i.e., ‘low’, ‘medium’ and ‘high’. The impact is calculated within the VLM by leveraging the O-RAN Domain Knowledge, which includes generic O-RAN WG11 specifications as well as domain-specific guidelines existing in the literature~\cite{baguer2024attacking}. Additionally, the output includes the \textit{thinking} process of the VLM, for both increasing the reasoning performance (i.e., applying Chain-of-Thought) and helping in the interpretation of the results (e.g., by a human operator or external RCA algorithms).\newline

\noindent \textbf{Deployment Options:} \tool{} is designed as a non-real time TSAD service operating over windows of approximately 200s. This latency is dominated by VLM inference and is therefore matched to the non-RT RIC control loop ($>$~1s) rather than to latency-critical control loops. In this sense, \tool{} is an anomaly-diagnosis layer for SMO-level reasoning, not a sub-second mitigation loop. In a real-world deployment, the pipeline is built as an rApp within the Non-RT RIC, ingesting cross-interface KPIs through the O1 interface and Virtual Event Streaming (VES) as the telemetry wrapper. The O-RAN Domain Knowledge may be maintained and updated within the rApp as specifications evolve. The emitted machine-readable outputs of anomalous events are then mapped onto standard SRE troubleshooting workflows~\cite{sre_book_troubleshooting}, which decompose incident response into problem reporting, triage, examination and diagnosis. \tool{} is designed to operate as the \textit{first} stage of this pipeline as the outputs serve as problem reports (localized intervals and responsible interfaces).

\section{Evaluation}
\label{sec:results}

% \begin{table}[t!]
% \centering
% \caption{Aggregated anomaly detection performance of \tool{} compared with state-of-the-art benchmarks}
% \label{tab:anomaly_performance}
% \begin{tabular*}{\columnwidth}{l@{\extracolsep{\fill}}ccc}
% \toprule
% \textbf{Method} &  \textbf{Recall} & \textbf{F1-Score} & \textbf{Accuracy} \\ 
% \midrule
% \texttt{MSCRED} & 0.128  & 0.187 & 0.103 \\
% \texttt{TAMA} & 0.315  & 0.429  & 0.275 \\
% \texttt{VLM4TS} & 0.296  & 0.408 & 0.258 \\
% \texttt{TSAD Agents} & 0.384  & 0.500  & 0.338 \\
% \texttt{DAST (w/o O-RAN Domain Know.)} & 0.502  & 0.652  & 0.495 \\
% \midrule
% \textbf{DAST} & \textbf{0.844}  & \textbf{0.910 } & \textbf{0.843 } \\ 
% \bottomrule
% \end{tabular*}
% \end{table}

\subsection{Testbed and Scenario}

\begin{figure}[t!]
  \centering
  \includegraphics[width=0.99\linewidth]{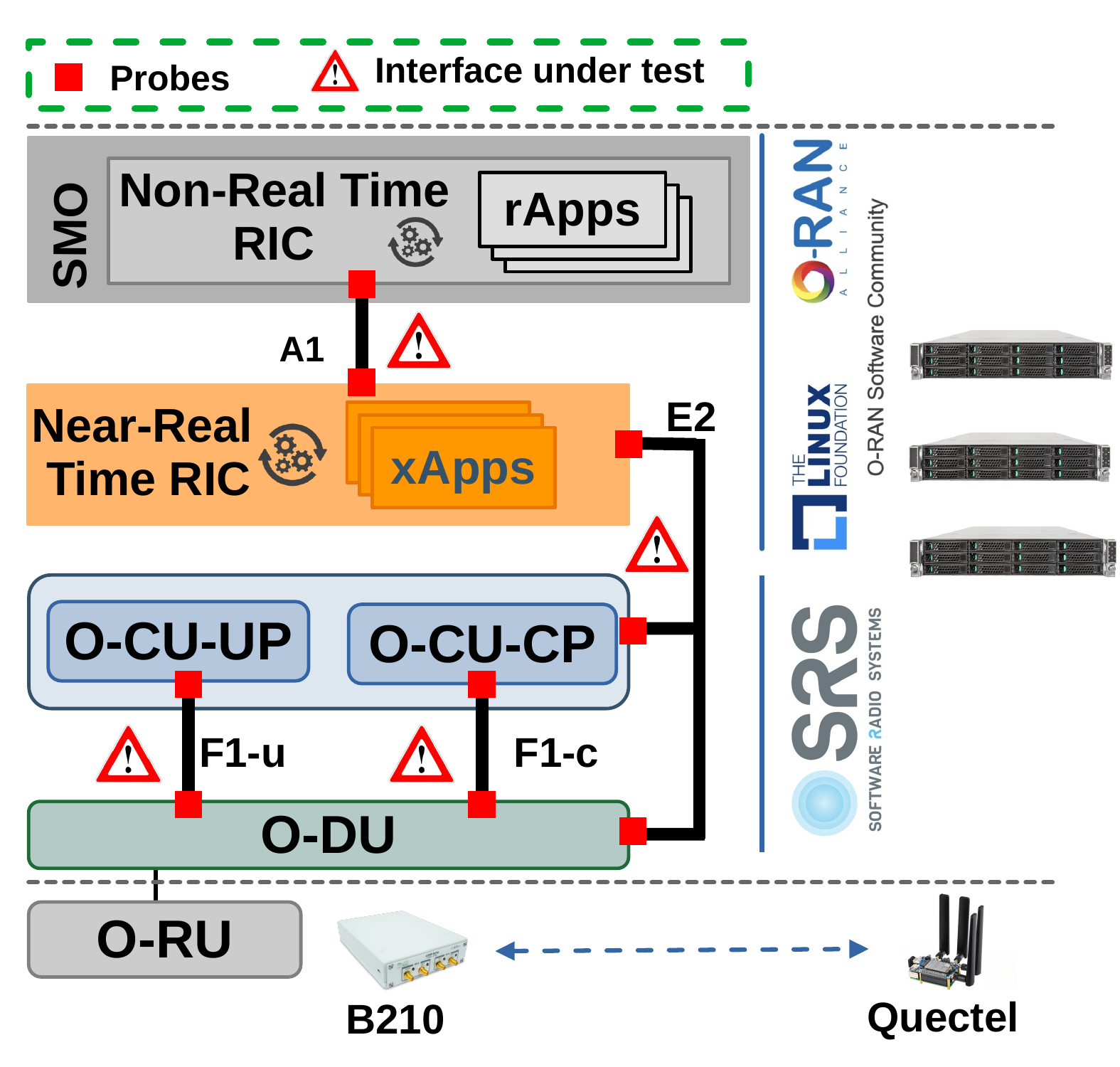}
  \caption{O-RAN testbed used for \tool{} evaluation.}
  \label{fig:dast_testbed}
\end{figure}

In order to evaluate the \tool{} framework, we deploy an experimental O-RAN testbed to generate real network traces with anomalies as shown in Fig.~\ref{fig:dast_testbed}. The O-RU, O-DU and O-CU components of the gNB are deployed using the srsRAN software stack\footnote{\url{https://www.srsran.com/5g}, Accessed: January 2026} implementing the functional split 8, while the near-RT RIC and non-RT RIC are built using the O-RAN SC projects\footnote{\url{https://o-ran-sc.org/}, Accessed: January 2026} and the core network with Open5GS\footnote{\url{https://open5gs.org/}, Accessed: January 2026}. We use Ettus USRP B210s for the 5G radio frontend and two Quectel RM520N-GL serving as UEs. 
To create a realistic network environment, the testbed implements FALCON~\cite{falkenberg2019falcon} to reproduce mobile traffic patterns captured in a real commercial network through MGEN. We selectively perform network performance degradation attacks on the F1-u, F1-c, A1, and E2 interfaces, increasing packet latency and packet loss with various degrees of impact (i.e., low, medium and high). 
\update{In total, we created a dataset of 1062 evaluations ($354$ traces x $3$ runs), distributed as follows: E2 (39.27\%), F1-u (35.03\%), A1 (12.71\%), F1-c (7.63\%), and baseline (5.37\%).} One trace typically lasts around 3 minutes, with the targeted network performance degradation attack randomly starting around the midpoint.  On average, typical packet delays range from 10 ms (low) to 5000 ms (high), while packet loss varies from 25\% (low) to 89\% (high). Fig.~\ref{fig:example} shows a visual example of one experiment's behavior.

Fig.~\ref{fig:dast_testbed} shows the probe location at these interfaces where network metrics were monitored.

\tool{} framework runs on a separate server with access to both the testbed and an Ollama instance running on a high-performance server (Intel Xeon Silver 32 CPU cores, 72 GB of RAM and two RTX Nvidia A5000 GPUs), where we make available the \textit{qwen3.6:35b} open-weight model\footnote{\update{released in April 2026, served using Ollama version 0.21.1 and with quantization Q4\_K\_M.}} used as both LLM and VLM.
\update{
For the inference settings, we relied on the model's internal defaults defined in its Ollama Modelfile. To accommodate the integration of multivariate time-series visual plots alongside textual O-RAN Domain Knowledge, the maximum context window size is set to 262144. The sampling parameters were set to a generation \texttt{temperature} of 1, \texttt{top-p} of 0.95, and \texttt{top-k} of 20, with a \texttt{presence-penalty} of 1.5 and \texttt{repeat-penalty} 1.0 (disabled). The maximum output length was left unconstrained to allow the model flexibility in its chain-of-thought, and we did not fix the random seed. The prompt design strictly follows a zero-shot template, concatenating the visual telemetry data, the O-RAN Domain Knowledge and a structured output (in JSON) instruction requiring the model to return the start and end timestamps of the anomalous period.}

% \begin{table}[t!]
% \centering
% \caption{Aggregated anomaly detection performance of \tool{} compared with state-of-the-art benchmarks}
% \label{tab:anomaly_performance}
% \begin{tabular*}{\columnwidth}{l@{\extracolsep{\fill}}cc}
% \toprule
% \textbf{Method} & \textbf{F1-Score} & \textbf{Accuracy} \\ 
% \midrule
% \texttt{MSCRED}~\cite{zhang2019mscred} & 0.187 & 0.103 \\
% \texttt{TAMA}~\cite{zhuang2024itthinkitsorted} & 0.429  & 0.275 \\
% \texttt{VLM4TS}~\cite{He_Alnegheimish_Reimherr_2026} & 0.408 & 0.258 \\
% \texttt{TSAD Agents}~\cite{xu2026can} & 0.500  & 0.338 \\
% \update{\texttt{\tool{} (w/o O-RAN Domain Knowledge)}} & 0.652  & 0.495 \\
% \midrule
% \textbf{\tool{}} & \textbf{0.910 } & \textbf{0.843 } \\ 
% \bottomrule
% \end{tabular*}
% \end{table}

\begin{table}[t!]
\centering
\caption{Aggregated anomaly detection performance of \tool{} compared with state-of-the-art benchmarks \update{(mean and 95\% CI over three runs)}}
\label{tab:anomaly_performance}
\begin{tabular*}{\columnwidth}{l@{\extracolsep{\fill}}cc}
\toprule
\textbf{Method} & \textbf{F1-Score} & \textbf{Accuracy} \\ 
\midrule
\texttt{MSCRED}~\cite{zhang2019mscred} & 0.247 $\pm$ 0.021 & 0.141 $\pm$ 0.014 \\
\texttt{TAMA}~\cite{zhuang2024itthinkitsorted} & 0.429 $\pm$ 0.020 & 0.275 $\pm$ 0.016 \\
\texttt{VLM4TS}~\cite{He_Alnegheimish_Reimherr_2026} & 0.408 $\pm$ 0.047 & 0.258 $\pm$ 0.049 \\
\texttt{TSAD Agents}~\cite{xu2026can} & 0.500 $\pm$ 0.009 & 0.338 $\pm$ 0.010 \\
\update{\texttt{\tool{} (no Domain Know.)}} &\update{ 0.652 $\pm$ 0.008 }& \update{0.495 $\pm$ 0.006} \\
\midrule
\textbf{\tool{}} & \textbf{0.911 $\pm$ 0.022} & \textbf{0.844 $\pm$ 0.036} \\
\bottomrule
\end{tabular*}
\end{table}

\begin{table*}[h]
\centering
\caption{Average Results per different O-RAN interfaces}
\label{tab:interface_results}
\resizebox{\textwidth}{!}{
\begin{tabular}{ll ccc ccc ccc ccc}
\toprule
\multirow{2}{*}{\textbf{Solution}} & \multirow{2}{*}{\textbf{Method}} & \multicolumn{3}{c}{\textbf{E2}} & \multicolumn{3}{c}{\textbf{F1-c}} & \multicolumn{3}{c}{\textbf{F1-u}} & \multicolumn{3}{c}{\textbf{A1}} \\
\cmidrule(lr){3-5} \cmidrule(lr){6-8} \cmidrule(lr){9-11} \cmidrule(lr){12-14}
 & & Precision & Recall & F1-Score & Precision & Recall & F1-Score & Precision & Recall & F1-Score & Precision & Recall & F1-Score \\
\midrule
\multirow{2}{*}{\texttt{MSCRED}~\cite{zhang2019mscred}} 
    %& Standard     & 0.000 & 0.000 & 0.000 & 0.000 & 0.000 & 0.000 & 0.560 & 0.226 & 0.322 & 0.789 & 0.333 & 0.469 \\
    %& Range-Wise   & 0.423 & 0.211 & 0.282 & 0.445 & 0.177 & 0.254 & 0.552 & 0.420 & 0.477 & 0.641 & 0.536 & 0.584 \\
    %& Standard     & 0.190 & 0.079 & 0.111 & 0.000 & 0.000 & 0.000 & 0.651 & 0.268 & 0.379 & 0.764 & 0.324 & 0.454 \\
    %& Range-Wise   & 0.451 & 0.205 & 0.281 & 0.438 & 0.140 & 0.210 & 0.681 & 0.486 & 0.566 & 0.708 & 0.599 & 0.648 \\
      & Standard     & 1.000 & 0.079 & 0.146 & 0.000 & 0.000 & 0.000 & 1.000 & 0.268 & 0.421 & 1.000 & 0.324 & 0.487 \\
      & Range-Wise   & 0.451 & 0.205 & 0.281 & 0.438 & 0.140 & 0.210 & 0.681 & 0.486 & 0.566 & 0.708 & 0.599 & 0.648 \\
\midrule
\multirow{2}{*}{\texttt{TAMA}~\cite{zhuang2024itthinkitsorted} } & Standard & 1.000 & 0.657 & 0.792 & 1.000 & 0.235 & 0.381 & 0.000 & 0.000 & 0.000 & 1.000 & 0.341 & 0.508 \\
 & Range-Wise & 0.946 & 0.802 & 0.868 & 0.964 & 0.415 & 0.579 & 0.950 &  0.317 &  0.475  & 0.957 & 0.581 &  0.723 \\
\midrule
\multirow{2}{*}{\texttt{VLM4TS}~\cite{He_Alnegheimish_Reimherr_2026} } & Standard & 1.000 & 0.381 & 0.549 & 1.000 & 0.209 & 0.346 & 1.000 & 0.363  & 0.525  & 1.000  & 0.211 & 0.349 \\
 & Range-Wise & 0.721  & 0.874  & 0.790  & 0.667 & 0.774 & 0.716 & 0.762 & 0.862  &  0.808 & 0.636 & 0.865 & 0.733 \\
\midrule
\multirow{2}{*}{\texttt{TSAD Agents}~\cite{xu2026can}} & Standard & 1.000 & 0.400 & 0.571 & 1.000 & 0.214 & 0.353 & 1.000 & 0.441 & 0.612 & 1.000 & 0.500 & 0.667 \\
 & Range-Wise & 0.846 & 0.705 & 0.769 & 0.729 & 0.739 & 0.734 & 0.794 & 0.801 & 0.797 & 0.750 & 0.743 & 0.746 \\
\midrule
\midrule
\multirow{2}{*}{\textbf{DAST}} & Standard & \textbf{1.000} & \textbf{0.875} & \textbf{0.933} & \textbf{1.000} & \textbf{0.679} & \textbf{0.808} & \textbf{1.000} & \textbf{0.828} & \textbf{0.905} & \textbf{1.000} & \textbf{0.889} & \textbf{0.941} \\
 & Range-Wise & 
 \textbf{0.990} & \textbf{0.846} & \textbf{0.912} & \textbf{0.995} & \textbf{0.703} & \textbf{0.822} & \textbf{0.978} & \textbf{0.854} & \textbf{0.901} & \textbf{0.987} & \textbf{0.868} & \textbf{0.924} \\
\bottomrule
\end{tabular}
}
\end{table*}

\subsection{Benchmarks}
We compare our approach against the following four benchmarks. We implemented approaches following classical TSAD methods and novel state-of-the-art solutions that rely on LLM/VLM frameworks specifically tailored for detecting anomalies in time series:

\begin{itemize}

\item  \texttt{MSCRED}~\cite{zhang2019mscred}: It uses a Multi-Scale Convolutional Recurrent Encoder-Decoder architecture to provide multivariate unsupervised anomaly detection. Requires training.

 \item \texttt{TAMA}~\cite{zhuang2024itthinkitsorted}: It is based on a 3-step (i.e., Reference Learning, Analyzing, and Self-reflection) multimodal VLM framework to detect anomalies in univariate timeseries, using labelled examples to learn normal data patterns. We extended the original framework to perform multivariate anomaly detection for a fair comparison.

 \item \texttt{VLM4TS}~\cite{He_Alnegheimish_Reimherr_2026}: This zero-shot framework is a combination of a Vision Encoder to detect local candidate anomalies in univariate time series and a VLM-based step to refine the preliminary detections with global timeseries knowledge. We extended the original framework to consider multivariate time series. 
       
 \item \texttt{TSAD Agents}~\cite{xu2026can}: This zero-shot solution is a multi-agent, multi-variate framework containing four agents that reason about anomaly types and adaptively route detection tasks to either traditional ML algorithms or VLMs using textual and visual inputs.

\end{itemize}

\update{Furthermore, to evaluate univariate baselines (\texttt{TAMA}, \texttt{VLM4TS}) fairly on our multivariate datasets, we utilized a minimal extension: they were supplied with the exact same domain context information and stacked, multi-channel visual plots provided to \tool{}, using identical VLM backbones and inference settings. }

\subsection{Results}
We evaluate \texttt{DAST} and the above mentioned benchmarks with two different sets of metrics. First, we use standard Recall, Precision, F1-Score and Accuracy, where a predicted interval counts as a True Positive only when it overlaps at least 70\% of the ground truth anomaly period while not exceeding its extent by more than 30\%. This regime is deliberately strict, since it rewards detections that pinpoint the anomaly's exact boundaries. 
However, this standard approach collapses each interval to a single binary hit, which can be misleading for anomalies of unequal duration and is susceptible to the well-documented Point-Adjustment illusion~\cite{Sarfraz2024}. We therefore additionally report Range-Wise Precision, Recall and F1-Score~\cite{tatbul2018precision}, which operate at the level of time ranges and assign proportional credit based on how much of each ground-truth interval is detected and how much of each prediction is correct.
% \update{We aggregate scores using a micro-averaging strategy at the run level. Instead of averaging the final percentages of individual experiments, the raw variables (TP, FP, FN, TN, and range-wise overlap sums) are calculated per experiment and pooled together across all experiments within a specific run. The final metrics are computed from these pooled totals. The values reported in the overall tables represent the mean of these run-level scores.
% % We don't show the confidence intervals due to lack of space and the variance being minimal.
% We report standard deviations only in Table~\ref{tab:anomaly_performance}, omitting them in subsequent results due to minimal variance.
% }
%\update{We aggregate scores using a micro-averaging strategy at the run level. Instead of averaging the final percentages for all individual experiments, the raw variables (TP, FP, FN, TN, and range-wise overlap sums) are calculated per experiment and averaged per run. The final metrics are reported in Table~\ref{tab:anomaly_performance} as mean and 95\% CI of these run-level scores.}
\update{We aggregate scores using a micro-averaging strategy at the run level. Instead of averaging the final percentages for all individual experiments, 
TP/FP/FN/TN and range-wise overlap sums are summed over all experiments of a run and the metrics are computed from these per-run totals. %Then, metrics are computed per run and 
Table~\ref{tab:anomaly_performance} reports the mean and 95\% CI over the three runs.}

\begin{figure*}[h]
  \centering
  \includegraphics[width=\linewidth]{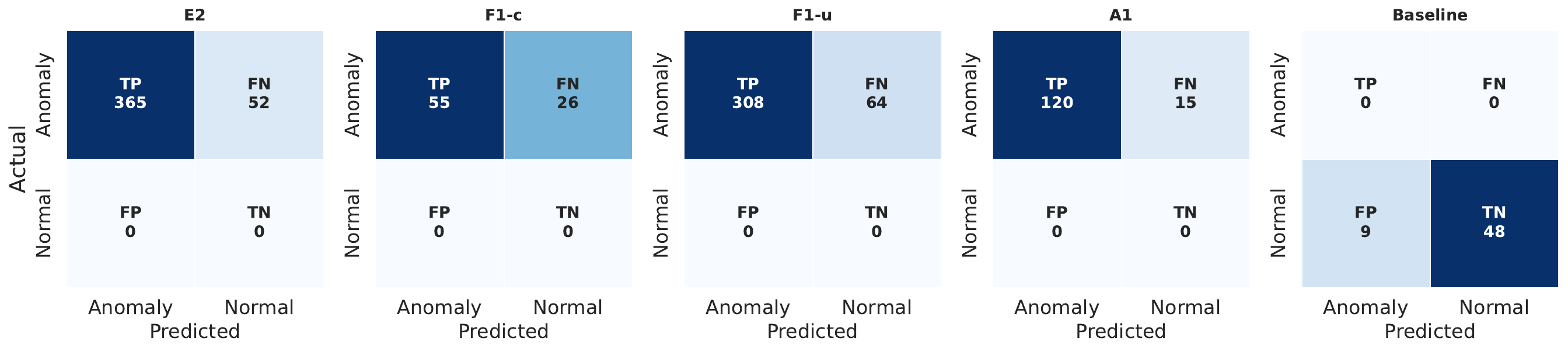}
  % \caption{\tool{} confusion matrices for anomaly detection.}
  \caption{Per-interface confusion matrices for \tool{} (E2, F1-c, F1-u, A1, and the non-anomalous baseline).}
  \label{fig:confusion}
\end{figure*}

% Table~\ref{tab:anomaly_performance} provides an aggregated overview of F1-Scores and Accuracy across all monitored interfaces \update{ (i.e., the 4 O-RAN interfaces plus the baseline), in order to measure \tool{}'s ability to distinguish between anomalous and normal states.}
Table~\ref{tab:anomaly_performance} provides an aggregated overview of F1-Scores and Accuracy across all \update{experiments (i.e., the ones with anomalies in the four O-RAN interfaces plus the non-anomalous baselines), in order to measure DAST's ability to distinguish between anomalous and normal states.}
\tool{} improves F1-Score by about 0.41 and Accuracy by 0.51 over the strongest baseline. \texttt{TAMA} and \texttt{VLM4TS} achieve worse \update{performance because 
% they were originally conceived for univariate detection and are not able to generalize to a multivariate O-RAN setting. 
loading many numerical plots into a single visual prompt induces hallucinations and dilutes the VLM's attention.}
On the other hand, multi-agent approaches (\texttt{TSAD Agents}, \tool{}) consistently dominate, confirming that the multivariate nature of O-RAN telemetry is best handled by multi-agent architecture, spreading detection tasks across specialised sub-agents rather than relying on a single foundation-model call. Still, the two approaches differ. While \texttt{TSAD Agents} is based on a general-purpose 4-stage setup that uses both VLMs and traditional ML algorithms, \tool{} uses a VLM $\rightarrow$ LLM $\rightarrow$ VLM pipeline.
In this sense, \tool{} achieves better performance due to its explicit integration of O-RAN Domain Knowledge, which narrows down the anomaly search space 
% \tocheck{and the use of temporally aligned multi-channel plots},
and helps to better capture cross-interface anomalies. Regarding \texttt{MSCRED}, despite being a well-established unsupervised multivariate detector, it performs the worst. This is because the highly stochastic profile of real O-RAN traces (from commercial base stations) creates an unsuitable scenario for reconstruction-based detectors that assume a stable baseline that can be learned offline. This result reinforces the case for zero-shot pipelines such as \tool{}, which inject O-RAN Domain Knowledge at inference time rather than depending on representative training traces.
\update{To further validate the impact of the O-RAN Domain Knowledge grounding, we conducted an ablation study (shown in Table~\ref{tab:anomaly_performance}). Removing the domain knowledge causes a severe performance drop, decreasing the F1-Score from $0.911$ to $0.652$ and Accuracy from $0.844$ to $0.495$. This confirms that integrating domain-specific grounding is a major contributor behind \tool{}'s superior anomaly detection capabilities.}

\begin{figure}[h!]
  \centering
  \includegraphics[width=0.99\linewidth]{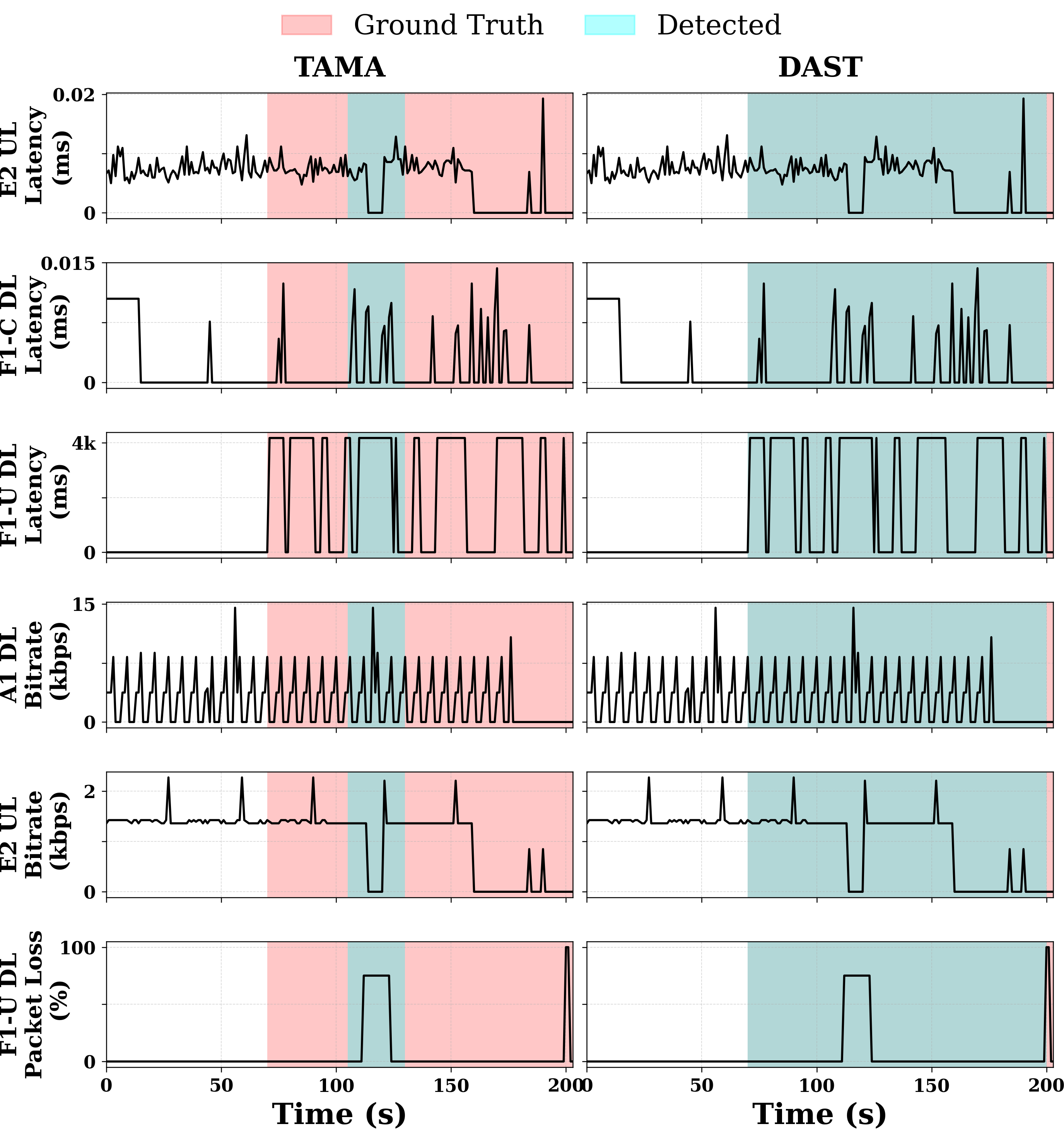}
  \caption{By leveraging O-RAN Domain Knowledge, \tool{} is able to realize that latency in F1-u is the actual anomaly, rather than sporadic packet loss in F1-u, which is actually the consequence.}
  \label{fig:example}
\end{figure}

Next, Table~\ref{tab:interface_results} shows the Precision, Recall and F1-Score for the interfaces considered in our testbed. \tool{} shows a superior performance compared to other benchmarks for all interfaces, with F1-c being the interface whose anomalies are most difficult to predict due to the inherent difficulty of isolating subtle anomalies on a low-traffic signalling control plane. For anomalies in other interfaces with periodic traffic such as E2 or with bulky stochastic traffic such as F1-u, detection and delimitation are effective.
On average, the Range-wise metric obtains higher values than the standard metric, since it calculates proportional overlaps and gives them partial credit, while standard has a strict threshold rule that penalizes more severely when not providing the exact overlap. While others can have a significant drop 
in terms of performance, \tool{} remains robust. This shows not only that \tool{} can detect anomaly intervals but that its detections precisely match the exact anomalous time interval.

\update{ It should be noted that the $0.000$ standard F1 scores registered by baselines like \texttt{MSCRED} and \texttt{TAMA} on specific interfaces are not indicative of configuration failures. Rather, they are artifacts of our strict event-level evaluation criteria ($\ge 70\%$ overlap, $\le 30\%$ over-extension). As demonstrated by their non-zero Range-Wise scores in Table~\ref{tab:interface_results}, these baselines do raise detections during anomalous windows, but their failure mode lies in imprecise temporal localization that misses the strict boundary thresholds. }

\update{Fig.~\ref{fig:confusion} reports per-anomalous-interface confusion matrices, which add further insights on Table~\ref{tab:interface_results} by exposing the structure of DAST's errors. The dominant pattern is asymmetric. On the four anomalous interfaces, DAST shows a false-negative rate of 0.12, 0.32, 0.17 and 0.11 respectively for the four interfaces.
F1-c shows the highest miss rate, in line with the data provided in Table~\ref{tab:interface_results}. On the 57 non-anomalous baseline experiments, DAST raises an alarm on 9 (15.8\% false-alarm rate), setting a non-zero floor on its false-alarm rate but well below the over-triggering behaviour of the considered benchmarks (see again Table~\ref{tab:anomaly_performance}). This rate is compatible with DAST's intended role as a problem-reporting stage that feeds human or automated triage rather than triggering mitigation directly~\cite{sre_book_troubleshooting}. Reducing this false-alarm floor is part of our future work.}
% \update{Finally, Fig.~\ref{fig:confusion} also shows a clear imbalance in the dataset, since only 57 out of 1062 experiments are non-anomalous. In order to provide a more robust Accuracy figure, we calculate Balanced Accuracy, that weights anomalous and non-anomalous experiments equally. However Balanced Accuracy (average between Recall + Specificity) results in 0.843, which for DAST is practically equivalent to the Accuracy figure provided in Table~\ref{tab:anomaly_performance}.}
    \update{Fig.~\ref{fig:confusion} also shows a clear imbalance in the dataset, since only 57 out of 1062 experiments are non-anomalous. This makes the Accuracy reported in Table~\ref{tab:anomaly_performance} to collapse mathematically into recall given the small proportion of baselines in the dataset. % (57 against 1062). %However, as we discuss later in this section, the Balanced Accuracy happens to be almost the same value.}
    To give a more robust Accuracy figure, here we provide Balanced Accuracy, which weights anomalous and non-anomalous experiments equally. The Balanced Accuracy (average of Recall and Specificity) results in 0.843, which for DAST happens to be practically equivalent to the Accuracy figure shown in Table~\ref{tab:anomaly_performance}.}
    
% Finally, beyond aggregate scores, Fig.~\ref{fig:example} illustrates why domain grounding matters in practice. In a scenario with high packet-delay on the F1-u, a purely visual detector such as TAMA is drawn to the most significant deviations (e.g., flags the packet loss region in F1-u). \tool{}, by contrast, cross-references the visual profile against the O-RAN domain knowledge, and therefore recognizes that the packet-loss spikes in F1-u are a downstream consequence, and the elevated and sustained F1-u latency is the actual root anomaly.
\update{Finally, beyond aggregate scores, Fig.~\ref{fig:example} qualitatively illustrates why domain grounding matters in practice. In a scenario with high packet-delay on the F1-u, a purely visual detector such as TAMA is drawn to the most significant deviations (e.g., flags the packet loss region in F1-u). \tool{}, by contrast, cross-references the visual profile against the O-RAN Domain Knowledge, helping to distinguish that the packet-loss spikes in F1-u are a consequence of the primary deviation found in the F1-u latency time series. }

\section{Future Directions}
\label{sec:future-directions}

% While \tool{} demonstrates that domain-grounded multi-agent reasoning is an effective basis for O-RAN anomaly detection, there are still open directions:\newline

% \noindent\textbf{Multi-vendor and zero-day validation}: Our evaluation uses a single open-source stack. A natural next step is to validate DAST across heterogeneous, multi-vendor O-RAN deployments and expand the performance degradation patterns considered in our evaluation. This will reinforce the results and substantiate the zero-shot generalization.

% \noindent \textbf{Closing the autonomous loop}: \tool{} produces reports and triage tags to be consumed downstream. A compelling extension is to close the loop by using this information to identify the root cause and generate policies to address the identified anomaly.

\update{While \tool{} demonstrates that domain-grounded multi-agent reasoning is an effective basis for O-RAN anomaly detection, there are still open directions:

\begin{itemize}
    \item \noindent\textbf{Multi-vendor and unseen-threat validation}: Our current evaluation uses a single open-source stack. A natural next step is to validate DAST across heterogeneous, multi-vendor O-RAN deployments and expand the performance degradation patterns considered in our evaluation. This will substantiate the pipeline's ability to generalize to unseen network states.
    \item \noindent \textbf{Root-cause and impact validation}: While \tool{} provides indicative impact ratings and qualitative insights into anomaly origins, validating its root-cause identification and impact estimation capabilities against standardized benchmarks remains a key future goal.
    \item \noindent \textbf{Closing the autonomous loop}: \tool{} produces reports and triage tags to be consumed downstream. A compelling extension is to close the loop by using this information to automatically generate policies to address the identified anomaly.
\end{itemize}
}

\section{Conclusion}
\label{conclusion}

Detecting performance degradation patterns in disaggregated O-RAN deployments is structurally hard given that ground truth baselines are scarce, distribution shift outpaces any feasible retraining cycle, and cross-interface cascade effects make per-interface detectors unreliable. \tool{} addresses all three by decomposing the detection across a zero-shot, three-stage VLM $\rightarrow$ LLM $\rightarrow$ VLM cascade in which each agent operates on the representation it is best suited to (visual profiling, textual reasoning and high-resolution visual identification). Evaluated on an open-source O-RAN testbed under representative latency and packet loss injection effects, \tool{} achieves an aggregated F1-Score of 0.911, outperforming both unsupervised multivariate detectors and recent LLM/VLM-based baselines. The broader implication is methodological. Our results suggest that anomaly detection in a disaggregated, multi-vendor RAN is better treated as multi-agent reasoning grounded in domain knowledge than as a single-model numerical-fitting problem, specifically gaining by injecting O-RAN semantics and from temporally aligned multi-channel plots that expose all anomalous behaviours across interfaces. For operators, this points to a practical observability primitive that can be maintained by editing a knowledge base rather than by collecting and labelling traces every time a vendor, slice, or performance degradation pattern changes. By emitting machine-readable problem reports, \tool{} is positioned not as a standalone classifier but as the entry point of an autonomous, SRE-style troubleshooting pipeline for the 6G compute continuum.

\section{Acknowledgement}
  
This work has been supported by the European Commission through grant no. SNS-JU-101139270 (ORIGAMI Project).

%%it does not count in the page count
 \bibliographystyle{IEEEtran}
 \bibliography{references}
 
  \begin{IEEEbiographynophoto}{} Francesco Spinelli is a Researcher at the i2CAT Foundation, Barcelona, Spain, working on GenAI for O-RAN. He received his Ph.D. from UC3M while at
  IMDEA Networks Institute, and was previously with Telecom Paris and a visiting researcher at UC Santa Cruz.
  \end{IEEEbiographynophoto}

  \begin{IEEEbiographynophoto}{} Esteban Municio is a Senior Researcher in the AI-driven Systems group at i2CAT and Adjunct Professor at UC3M. He received his Ph.D. from the University
  of Antwerp (imec-IDLab) in 2020. His interests include AI-driven open networks, Open RAN, TSN, Industrial IoT and NTN.
  \end{IEEEbiographynophoto}

  \begin{IEEEbiographynophoto}{} Pau Baguer is a Junior Researcher at i2CAT and is pursuing an M.Sc. in Artificial Intelligence at UPC. His work focuses on integrating intelligent
  systems into Open RAN, including O-RAN-based indoor localization with smart surfaces and Open 5G/6G network optimization for BVLOS UAV operations.
  \end{IEEEbiographynophoto}

  \begin{IEEEbiographynophoto}{} Gines Garcia-Aviles received his M.Sc. (2018) and Ph.D. (cum laude, 2021) in Telematics Engineering from UC3M, while at IMDEA Networks Institute. Since
  2021 he has been a post-doctoral researcher at i2CAT. His interests include NFV, network slicing, resource orchestration and AI.
  \end{IEEEbiographynophoto}

  \begin{IEEEbiographynophoto}{} Xavier Costa-P\'erez is an ICREA Research Professor, Scientific Director at i2CAT and head of ASTRO R\&D at NEC Laboratories Europe. He received his M.Sc. and Ph.D. from UPC, has contributed to 3GPP, O-RAN, IEEE 802.11 and IETF, and holds about 100 granted patents.
  \end{IEEEbiographynophoto}

\end{document}